# Predicting Adversary Lateral Movement Patterns with Deep Learning

Nathan Danneman[*]
Data Machines
James Hyde
Data Machines

April 23, 2021

## Abstract

This paper develops a predictive model for which host, in an enterprise network, an adversary is likely to compromise next in the course of a campaign. Such a model might support dynamic monitoring or defenses. We generate data for this model using simulated networks, with hosts, users, and adversaries as first-class entities. We demonstrate the predictive accuracy of the model on out-of-sample simulated data, and validate the findings against data captured from a Red Team event on a live enterprise network.

## 1 Introduction

Being able to predict what hosts an adversary will target *next* during a cyber campaign would be invaluable. This knowledge would enable targeted hunting, with defenders being able to focus their attention on hosts that are likely to be compromised next. Understanding which hosts are most likely to be compromised next also supports *dynamic* defenses and monitoring. This research grew out of DARPA's Cyber Hunting at Scale (CHASE) Program[1], which takes as given the fact that enterprise systems produce more cyber-relevant data than they can analyze. Armed with a sense of where in an enterprise an adversary might pivot next, defenders could target onerous user-facing security policies and increased monitoring/collection intelligently, rather than increasing blanket defense and collection postures across the enterprise.

The ideal data to support a predictive model for which host an adversary will compromise next, conditioned on defender knowledge of an enterprise and a list of hosts an adversary is currently thought to have compromised, might come from watching hundreds of adversaries attempt to traverse thousands of real networks. Clearly, such a dataset does not exist. This paper attacks that data problem by simulating networks and attack campaigns to support such

---

[*]nathan.danneman@datamachines.io

[1]Distribution Statement "A" (Approved for Public Release, Distribution Unlimited)

This research was developed with funding from the Defense Advanced Research Projects Agency (DARPA). The views, opinions, and/or findings expressed are those of the author(s) and should not be interpreted as representing the official views or policies of the Department of Defense or the U.S. Government.

predictions. Section 2 describes network simulations, including simulated adversaries with goals, realistic reconnaissance methods, and heuristic patterns for traversing networks. From these simulations , we generate data on adversaries attacking networks, and emit that data for consumption in the training phase of a deep neural network which predicts which host an adversary will move to next; see Section 3. The potency of this setup is tested on out-of-sample simulated data, as well as on real world data from a Red Team exercise undertaken on a live network.

Nuanced simulations, e.g. those generated by ns3 (Riley and Henderson 2010), are not the focus of this paper. Rather, the focus is on adversary actions at the host level; realistic network traffic is not needed to infer host sequences. However, this paper requires more nuance than a mere network topology, e.g. from TopoGen (Laurito et al. 2017). There is a wide-ranging set of literature that tries to predict adversary actions. Some focuses on host-level activities, e.g. Colbaugh and Glass (2012), while others leverage game theory to try to make inferences about network-level actions, e.g. Horák, Zhu and Bošanskỳ (2017). This paper diverges from those bodies of research by focusing at the network level, and empirically modeling a simulated adversary across networks, rather than assuming adversary value functions within networks.

## 2  Simulations

We simulate networks, adversaries, and campaigns against individual networks. Detailed descriptions of these simulations follow, as they inform the predictive model as well as the methods for linking the simulation study to real world data. The simulated networks feature users and hosts as key components. Users are represented as having integer-valued privilege levels, access to certain subnets, and have a history of having accessed a particular set of hosts. Hosts form the other portion of a bipartite graph, opposite of users. Hosts are represented as existing on a single subnet, require an integer-valued privilege level to access, and have a history of accesses by a particular set of users. In these simulations, hosts also probabilistically contain exploits. One exploit allows an adversary to increase the privilege level of any user that has access to that host. A second class allows an adversary to access the credentials of any user that is in the recent users list on that host. Substantively, these simple exploit classes stand in for service-specific vulnerabilities that allow for privilege escalation and credential harvesting, respectively.

This paper simulates adversaries that are *goal oriented*, operationalized as attempting to reach a particular host (or type of host) on the network they have compromised. This assumption might prove limiting – clearly, some real-world adversaries are opportunistic rather than goal-directed. However, over short time horizons, both opportunistic and goal-directed adversaries are subject to similar feasibility constraints. Our simulated adversaries, like human ones, act with a weighted set of heuristics when attempting to traverse a network, locate exploitable vulnerabilities, and move towards their intended target. The heuristics include actions such as: "access higher-privilege hosts and more capable users when possible," "attempt to access users or hosts on previously unexplored subnets," and "escalate privileges of users whenever possible." When these heuristics are infeasible, adversaries randomly search hosts, looking for exploitable vulnerabilities and routes to their target.

In all simulations, adversaries begin on a host, with credentials for a particular user (e.g. as if from phishing), and with a goal host. The adversary attempts to traverse through the network until it either reaches its goal or finds itself blocked – that is, with no access to credentials for hosts or users it has not already evaluated for routes to its goal or exploitable vulnerabilities.

## 3  Model and Results

Data is simulated from tens of thousands of episodes of adversaries attempting to reach a randomized goal, from a randomized starting user-host combination, on a randomly simulated network. This data includes metadata about the users and hosts an adversary visits – their subnet, privilege level, type, etc. The data is ordered based on the order in which the adversary reaches each host. These structured simulation outputs become training data for a feed-forward neural network that predicts the metadata associated with which host will be accessed next. The predictors are one-hot encoded state variables characterizing the metadata associated with all previously exploited hosts and users – their subnets, privilege levels, types, and combinations of these factors. These predictors are used to predict the likelihood of the next visit by the adversary to each possible host machine not yet visited. We found a four layer model with batches from mixtures of different simulation runs to converge strongly and give the best predictive performance across a random architecture search.

At inference time, our model generates predictions over each possible subsequent host conditioned on the history of previously exploited hosts. This list of predicted probabilities is then sorted. This paper's goal is not to predict *exactly* which host will be exploited next, as there are frequently dozens or hundreds with the same metadata profiles (e.g. any other workstation in an enclave for which an adversary has permissions from a valid user). Rather, we desire that the actual next host visited is frequently towards the top of a sorted list of predicted probabilities of next hosts. Figure 1 displays the percentile of the predicted probability of the *actual* next host after a particular number of hosts has been compromised. The horizontal black line depicts chance guessing. The points are medians, with 80% confidence intervals. Blue depicts networks with around 200 hosts and users. The red marks depict networks with around 1,000 hosts and users. The green marks depict the model's ability to predict the next host if user-level information is not known by defenders. Overall, this simple predictive model over host and user metadata is potent: the median percentile of the predicted probability of the actual next host across simulations is around 75. Substantively, this means that the model typically puts the correct next host in the top quartile its predictions – a strong enough predictor to be actionable, e.g. by increasing monitoring on only 25% of hosts with high likelihood of that monitoring capturing an adversary's next lateral movement.

We validated the efficacy of this predictive model against *actual* adversaries attacking *actual* networks using data from a Red Team exercise. Red Team notes from a Cobalt Strike (Seazzu 2016) report detailed which users and hosts were compromised at each sequence of the attack. For convenience, Kerberos data, hostname, and username patterns from netflow logs were used to marry the real world network data with the representations native to the simulation. The predicted probability of the second compromised host, conditional on seeing the first host compromised by the Red Team was poor (41st percentile). This was due to the fact that that Red Team was white carded onto a workstation with no recent Kerberos history, and onto a fake/new user. The percentile of the predicted probability the model gave for the third compromised host, conditioned on observing the prior two, was 69. This is in line with the simulation findings with observable information on which hosts and users have been compromised. Over-all, this seems a moderate success; defensive actions based on this model could target around a third of hosts, rather than moving all hosts to an increased security posture, and have a high likelihood of increased visibility on hosts an adversary moves to next.

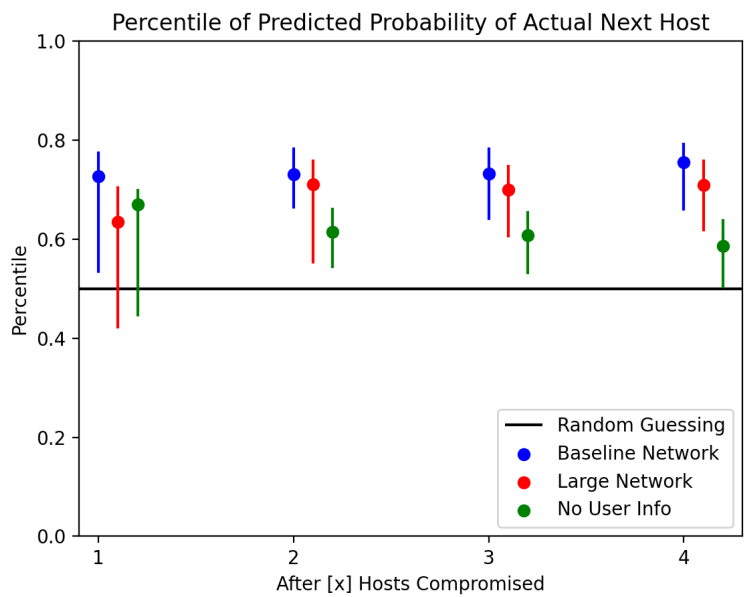

Figure 1

## 4 Conclusion

Overall, this work breaks new ground in using simulation studies and predictive models to make inferences about an adversary's likely next target in a large enterprise network. We show that adversary next actions can be predicted with sufficient fidelity to support dynamic monitoring or defenses in out-of-sample prediction tasks. Further, the usefulness of this technique was validated against real network data from a Red Team activity on a live network.

This work serves as a strong first step towards predicting subsequent adversary actions. Future work will expand on this effort in several directions. Foremost, more detailed simulations will be undertaken with more realistic role-based access controls and service-level exploit structures. Additionally, more potent predictive model classes will be employed, likely ones that leverage the graph structure of network data, as well as those that explicitly rank unordered sets of potential next hosts (e.g. pointer nets). Finally, more attention will be given to programmatically linking real network data to simulation structures, to build out a capability that is usable by defenders.